# General PFG signal attenuation expressions for anisotropic anomalous diffusion by modified-Bloch equations

Guoxing Lin*

Carlson School of Chemistry and Biochemistry, Clark University, Worcester, MA 01610, USA

**Abstract**

Anomalous diffusion exists widely in polymer and biological systems. Pulsed-field gradient (PFG) anomalous diffusion is complicated, especially in the anisotropic case where limited research has been reported. An general PFG signal attenuation expression, including the finite gradient pulse (FGPW) effect for free general anisotropic fractional diffusion { $0 < \alpha, \beta \leq 2$ } based on the fractional derivative, has not been obtained, where $\alpha$ and $\beta$ are time and space derivative orders. It is essential to derive a general PFG signal attenuation expression including the FGPW effect for PFG anisotropic anomalous diffusion research. In this paper, two recently developed modified-Bloch equations, the fractal differential modified-Bloch equation and the fractional integral modified-Bloch equation, were extended to obtain general PFG signal attenuation expressions for anisotropic anomalous diffusion. Various cases of PFG anisotropic anomalous diffusion were investigated, including coupled and uncoupled anisotropic anomalous diffusion. The continuous-time random walk (CTRW) simulation was also carried out to support the theoretical results. There is good agreement between the theory and the CTRW simulation. The obtained signal attenuation expressions and the three-dimensional fractional modified-Bloch equations are important for analyzing PFG anisotropic anomalous diffusion in NMR and MRI.

## 1. Introduction

Anomalous diffusion [1,2,3] has been investigated in many systems such as polymer or biological systems [4], porous materials [5,6], fractal geometries [7], micelle systems [8] and other systems [9,10]. The Anomalous diffusion has time derivative order $\alpha$ and space derivative order $\beta$, $0 < \alpha, \beta \leq 2$. When $\alpha = 1, \beta = 2$, the anomalous diffusion reduces to normal diffusion. Unlike normal diffusion, anomalous diffusion has a non-Gaussian propagator, and its mean $\beta$-th power of displacement is not linearly proportional to its diffusion time. Anomalous diffusion could be modeled by time-space fractional diffusion equations based on fractal derivative (see Appendix A) model and fractional derivative (see Appendix B) model [1,2,11,12,13,14,15,16,17,18,19]. Due to the non-Gaussian characteristic features of anomalous diffusion, it is difficult to investigate pulsed field gradient (PFG) anomalous diffusion. The PFG technique has been a powerful tool in studying normal diffusion [20,21,22,23,24], but many PFG normal diffusion theories may not be directly applicable to the investigation of anomalous diffusion without modification. There have been many efforts devoted to studying PFG anomalous diffusion [18,19,25,26,27,28,29,30,31,32,33,34,35,36], but most of them are related to isotropic anomalous diffusion. The theoretical research of PFG anisotropic anomalous diffusion is very limited [37,38,39].

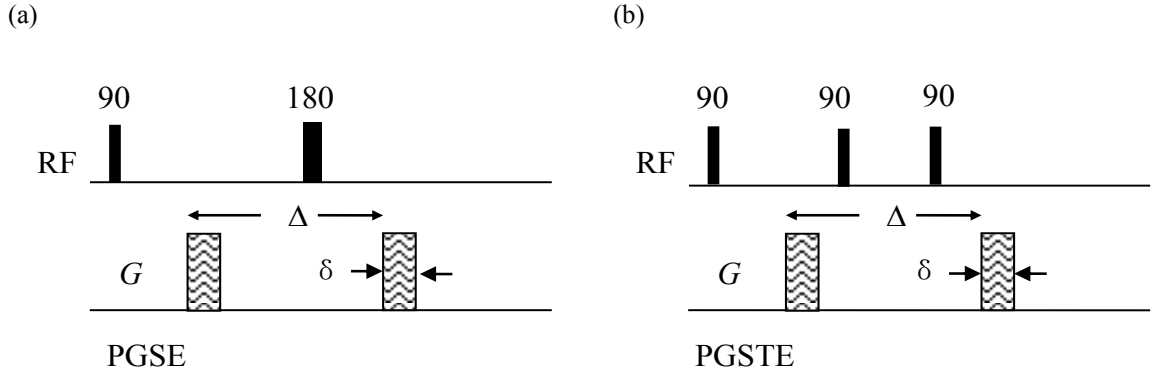

**Fig. 1** Two typical PFG pulses: (a) PGSE pulse sequences, (b) PGSTE pulse sequence. The gradient pulse width is $\delta$, and the diffusion delay is $\Delta$. In PGSTE pulse sequence, the magnetization is transferred to the Z direction by the second radio frequency (RF) pulse to eliminate the $T_2$ relaxation effect, which has the advantage over the PGSE pulse sequence.

Anisotropic diffusion behaviors could exist in many materials [40,41,42,43,44,45] such as stretched or compressed polymers, block polymers, liquid crystals, muscle cells, brain tissues and so on. In these anisotropic systems, the diffusion may belong to anisotropic anomalous diffusion. Compared to PFG anisotropic normal diffusion, the PFG anisotropic anomalous diffusion is much more complicated. First, the PFG signal attenuation in anomalous diffusion may be a stretched exponential function based attenuation from the fractal derivative [11-14] or a Mittag-Leffler function (MLF) based attenuation from the fractional derivative [3,15-17]. When $|x+y+z|$ is not small, the MLF function has $E_{\alpha,1}(-x-y-z) \neq E_{\alpha,1}(-x)E_{\alpha,1}(-y)E_{\alpha,1}(-z)$, which is different from the $\exp(-x-y-z) = \exp(-x)\exp(-y)\exp(-z)$ in anisotropic normal diffusion. Second, the characteristic feature of anisotropic normal diffusion is that it has three different diffusion coefficient components in axes $x', y', z'$ of the principal axis frame. While, the anisotropic anomalous diffusion have three sets of parameters including $D_{fi'i'}$, $\alpha_{i'}$ and $\beta_{i'}$, where $i' = x', y', z'$ and $D_{fi'i'}$ is the anomalous diffusion coefficient along the $i'$ axis. In addition, the anomalous diffusion along the three principal axes may be correlated (referred to as coupled diffusion) or non-correlated (referred to as uncoupled diffusion). Currently, for time fractional diffusion and general fractional diffusion, the theoretical results of PFG anisotropic anomalous diffusion based on the modified Bloch equation method in Ref. [37] only include the attenuation of the first gradient pulse, $0 < t \leq \delta$. In typical pulsed gradient spin echo (PGSE) or pulsed gradient stimulated echo (PGSTE) experiments as shown in Fig. 1, the PFG signal attenuation takes place in all three periods: $0 < t \leq \delta$, $\delta < t \leq \Delta$, $\Delta < t \leq \Delta + \delta$. Thus, it is important to derive general PFG signal attenuation expressions for the studying of PFG anisotropic anomalous diffusion.

This paper has three goals. First, to extend the recently developed fractal differential and fractional integral modified-Bloch equations [46] to analyze the three-dimensional anisotropic anomalous diffusion, which is important to study three-dimensional PFG anomalous diffusion. Various groups have proposed different modified Bloch equations for anomalous diffusion, which combine the fractional diffusion with precession and relaxation in significantly different manners [19,37,46]. The modified-Bloch equations proposed in Ref. [46] were employed here based on the following considerations: they provide the signal attenuation for the whole gradient pulse sequence, and their results can be reduced to those obtained by the short gradient pulse (SGP) approximation [46]. Additionally, when the fractional diffusion becomes very slow or eventually there is no fractional diffusion, the modified Bloch equations in Ref. [46] reduce to the traditional linear Lamour precession equation. Moreover, the numerical evaluation of the PFG signal attenuation based on this modified-Bloch equation is not difficult. The second goal of this paper is to obtain the general PFG signal attenuation expressions for anisotropic anomalous diffusion by these modified-Bloch equations. Third, to help us understand the PFG anomalous diffusion and to test the results from the phenomenological modified-Bloch equation, the continuous-time random walk (CTRW) simulation [47] is performed. Random walk simulation is a stochastic jump process, which is a powerful numerical method used to model normal and anomalous diffusion in physics, chemistry and many other disciplines [1,2,47]. The random walk can conveniently simulate the PFG signal attenuation, which can be used to test mathematically tractable systems and help us understand complicated systems that may be mathematically intractable in PFG normal diffusion and PFG anomalous diffusion. The CTRW method used in this paper has been used in references [33] and [36] to simulate PFG signal attenuation in free and restricted anomalous diffusions. This CTRW simulation method is based on two different models. The continuous waiting time and jump length are based on the CTRW model suggested in reference [47] whose applications are used to model anomalous diffusion in physics and economics. While the phase change and the PFG signal attenuation are simulated based on the algorithm from the lattice model developed in references [48,49], which has been used to model PFG diffusion in polymer systems [50]. Both the anisotropic diffusions with correlated or non-correlated $x', y', z'$ direction diffusions will be simulated. The simulation results agree perfectly with the results from the fractional integral modified Bloch equation.

All the theoretical results obtained here can be reduced to the reported anisotropic normal diffusion when $\alpha = 1$ and $\beta = 2$ [23,24], and isotropic anomalous diffusion results when the three parameters $D_{f'i'}$, $\alpha_{i'}$ and $\beta_{i'}$ are the same in different axis directions. Additionally, different numerical approaches can be used to calculate the theoretical PFG signal attenuation. In particular, for the fractional derivative, the PFG signal attenuation can be numerically evaluated either by the Adomian decomposition method [51,52,53,54,55] or a direct integration method (see Appendix C) that is proposed in this paper. Although the numerical values obtained from these two methods are identical, the calculation speed of the direct integration method is drastically faster than that of the Adomian decomposition method. The results here help us to understand PFG anisotropic anomalous diffusion and provide a valuable formalism for PFG anisotropic anomalous diffusion.

**2. Theory**

In PFG anisotropic diffusion, the effective gradient $\mathbf{g}'$ in the principal axis frame can be obtained by

$$\mathbf{g}' = R\mathbf{g}, \tag{1}$$

where $\mathbf{g}$ is the gradient vector in the observational reference frame, and $R$ is the rotation matrix that rotates a vector or tensor from the observational reference frame to the principal axis frame. When $\alpha_{x'} = \alpha_{y'} = \alpha_{z'}$, $\beta_{x'} = \beta_{y'} = \beta_{z'}$, the diffusion tensor can be transformed by

$$\mathbf{D_f} = \tilde{R} \cdot \mathbf{D'_f} \cdot R, \tag{2}$$

where $\mathbf{D_f}$ and $\mathbf{D'_f}$ are the diffusion tensors in the observational reference frame and the principal axis frame, respectively. In PFG experiments, the accumulating phase $\phi$ of a spin is [24,26,23]

$$\phi = -\int_0^t \gamma \mathbf{g}(t') \cdot \mathbf{r}(t') dt' = -\int_0^t \gamma \mathbf{g}'(t') \cdot \mathbf{r}'(t') dt', \tag{3}$$

where $\gamma$ is the gyromagnetic ratio, $\mathbf{r}(t)$ and $\mathbf{r}'(t)$ are the time-dependent position vectors in the observational reference frame and the principal axis frame, respectively. Averaging over all possible $\phi$ leads to signal attenuation [24,26]

$$A(t) = S(t)/S(0) = \int_{-\infty}^{\infty} P(\phi,t) \exp(-i\phi) d\phi, \tag{4}$$

where $A(t)$ is the signal attenuation, $S(0)$ and $S(t)$ are the signal amplitudes at the beginning of the first gradient pulse and at the end of the last gradient pulse, respectively, and $P(\phi,t)$ is the probability distribution function of the accumulating phase. For simplicity, $S(0)$ will be set as 1, then $A(t) = S(t)$. $S(t)$ will be used to denote the signal attenuation in the rest of this paper. As $S(t)/S(0) = \int_{-\infty}^{\infty} P(\phi,t) \exp(-i\phi) d\phi \leq 1$ is always true, in general, the signal intensity affected by diffusion is attenuated, which can be monitored in PFG experiments regardless of whether the diffusion is normal or anomalous. From Eqs. (3) and (4), the PFG signal attenuation expression can be analyzed directly inside the principal axis frame.

In anisotropic diffusion, the diffusions in the three principal axes are coupled or uncoupled, which can be seen from the continuous-time random walk [2,47]. In CTRW simulation, for coupled diffusion, the change of each random jump belonging to one of the three axes is proportional to certain ratios. For a simple case, the coupled diffusion may have $\alpha_{x'} = \alpha_{y'} = \alpha_{z'}$, and probability of jumping in a given direction ratio equals 1/3. It is still possible, that the coupled diffusion has different ($D_{f_{i'i'}}, \alpha_{i'}, \beta_{i'}$), $i' = x', y', z'$ on each axis, but the total jump time $t_{i'}$ of each axis is different. For an uncoupled diffusion, the diffusion along each axis direction is independent of the diffusion along other axes directions. For such an uncoupled diffusion, the parameter set ($D_{f_{i'i'}}, \alpha_{i'}, \beta_{i'}$), $i' = x', y', z'$ for three directions could be different. Such an uncoupled diffusion can be treated as a combination of three one-dimensional diffusions.

2.1 Coupled anisotropic diffusion

We first study the simple coupled anisotropic anomalous diffusion that has the same $\alpha_{x'} = \alpha_{y'} = \alpha_{z'} = \alpha$, but may have different $D_{f_{i'i'}}, \beta_{i'}$ on three principal axes, $x', y', z'$.

2.1.1 Fractional differential modified-equation based on the fractal derivative

In Ref. [46], the one-dimensional fractional differential modified Bloch equation based on the fractal derivative model is

$$\frac{\partial M(\mathbf{z},t)}{\partial t} = \alpha t^{\alpha-1} D_{f_1} \frac{\partial}{\partial z^\beta} M(\mathbf{z},t) - i\gamma \mathbf{g}(t)\cdot \mathbf{z} M(\mathbf{z},t) - \frac{M(\mathbf{z},t)}{T_2}, \qquad (5)$$

where $M(\mathbf{z},t)$ is the spin magnetization in the observational reference frame, $\mathbf{z}$ is the position and $T_2$ is the spin-spin relaxation time. The one-dimensional modified-Bloch equation (5) can be extended to three dimensions. The three-dimensional fractional differential modified Bloch equation based on the fractal derivative for isotropic diffusion will be

$$\frac{\partial}{\partial t} M(\mathbf{r},t) = \alpha t^{\alpha-1} D_{f_1} \nabla_{f_1}^\beta M(\mathbf{r},t) - i\gamma (\mathbf{g}(t)\cdot \mathbf{r}) M(\mathbf{r},t) - \frac{M(\mathbf{r},t)}{T_2}, \qquad (6)$$

where $\nabla_{f_1}^\beta = \frac{\partial}{\partial x^{\beta/2}}\left(\frac{\partial}{\partial x^{\beta/2}}\right) + \frac{\partial}{\partial y^{\beta/2}}\left(\frac{\partial}{\partial y^{\beta/2}}\right) + \frac{\partial}{\partial z^{\beta/2}}\left(\frac{\partial}{\partial z^{\beta/2}}\right)$. While, for anisotropic diffusion, the modified Bloch equation in the principal axis frame can be written as

$$\frac{\partial M'(\mathbf{r}',t)}{\partial t^\alpha} = \left[ D'_{f_1 x'x'} \frac{\partial}{\partial x'^{\beta_{x'}/2}}\left(\frac{\partial}{\partial x'^{\beta_{x'}/2}}\right) + D'_{f_1 y'y'} \frac{\partial}{\partial y'^{\beta_{y'}/2}}\left(\frac{\partial}{\partial y'^{\beta_{y'}/2}}\right) + D'_{f_1 z'z'} \frac{\partial}{\partial z'^{\beta_{z'}/2}}\left(\frac{\partial}{\partial z'^{\beta_{z'}/2}}\right) \right] \times$$
$$\alpha t^{\alpha-1} M'(\mathbf{r}',t) - i\gamma(\mathbf{g}'(t)\cdot \mathbf{r}') M(\mathbf{r}',t) - \frac{M'(\mathbf{r}',t)}{T_2}, \qquad (7)$$

where $M'(\mathbf{r}',t)$ is the spin magnetization in the principal axis frame, $\mathbf{g}'$ is the effective gradient in the principal axis frame, $D'_{f_1 i'i'}$, $i' = x',y',z'$ is the component of fractional diffusion coefficient tensor $\mathbf{D}'_{f_1}$ in the principal axis frame, and the units of $D'_{f_1 i'i'}$ are $m^{\beta_{i'}}/s^{\alpha_{i'}}$. When $D'_{f_1 x'x'} = D'_{f_1 y'y'} = D'_{f_1 z'z'} = D_{f_1}$ and $\beta_{x'} = \beta_{y'} = \beta_{z'} = \beta$, the anisotropic diffusion reduces to an isotropic diffusion.

For a system with homogeneous spin density, the magnetization in the principal axis frame can be described as $M'(\mathbf{r}',t) = S(t)\exp(-i\mathbf{K}'(t)\cdot \mathbf{r}')\exp(-\frac{t}{T_2})$ [37,46], where $K'_i(t') = \int_0^{t'} \gamma g'_i(t'')dt''$ is the wavenumber in the principal axis frame.

Substituting $M'(\mathbf{r}',t) = S(t)\exp(-i\mathbf{K}'(t)\cdot \mathbf{r}')\exp(-\frac{t}{T_2})$ into Eq. (7), we get

$$\frac{\partial S(t)}{\partial t} = -\alpha t^{\alpha-1}\left[\sum_{i'=x',y',z'} D'_{f_1 i'i'} K'^{\beta_{i'}}_{i'}(t)\right] S(t) - \frac{t}{T_2} S(t). \qquad (8)$$

The solution of the Eq. (8) is

$$A(t) = \exp(-\frac{t}{T_2}) \exp\left[-\sum_{i'=x',y',z'} \int_0^t D'_{f_1 i'i'} K'^{\beta_{i'}}_{i'}(t) dt^\alpha\right] = \exp(-\frac{t}{T_2}) \exp(-\mathbf{b}'_f : \mathbf{D}'_{f_1}). \qquad (9)$$

where

$$\mathbf{b}'_f = \int_0^t \left[\widetilde{\mathbf{K}}'_f \cdot R\right] \otimes \left[\widetilde{R}\cdot \mathbf{K}'_f\right] \alpha t'^{\alpha-1} dt', \qquad (10a)$$

and $\mathbf{K}'^{\beta_i/2}_f$ is defined as

$$\mathbf{K'}^{\beta_{i'}/2}{}_f = \begin{pmatrix} K'^{\beta_{x'}/2}_{x'}\mathbf{i}_{x'} \\ K'^{\beta_{y'}/2}_{y'}\mathbf{i}_{y'} \\ K'^{\beta_{z'}/2}_{z'}\mathbf{i}_{z'} \end{pmatrix} = \begin{pmatrix} \left(\sum_{j=x,y,z} R_{xj} K_j\right)^{\beta_{x'}/2} \mathbf{i}_{x'} \\ \left(\sum_{j=x,y,z} R_{yj} K_j\right)^{\beta_{y'}/2} \mathbf{i}_{y'} \\ \left(\sum_{j=x,y,z} R_{zj} K_j\right)^{\beta_{z'}/2} \mathbf{i}_{z'} \end{pmatrix}. \quad (10b)$$

For simplicity's sake, all theoretical calculations and the CTRW simulations in this paper are performed in the principle axis frame. In real application, $\mathbf{g}'$ and $\mathbf{K'}^{\beta_i/2}_f$ can be obtained by Eq. (1) and Eq. (10b) respectively.

2.1.2 Fractional integral modified-Bloch equation based on the fractional derivative

The one-dimensional modified-Bloch equation for PFG fractional diffusion based on the fractional derivative is an integral type equation [46]

$$M(\mathbf{z},t) = \sum_{k=0}^{m-1} M^{(k)}(\mathbf{z},0^+)\frac{t^k}{k!} + \int_0^t \left\{ \left( \frac{(t-\tau)^{\alpha-1}}{\Gamma(\alpha)} D_{f_2} \frac{\partial^\beta}{\partial|z|^\beta} - i\gamma \mathbf{g}(t)\cdot\mathbf{z} - \frac{1}{T_2} \right) M(\mathbf{z},t) \right\} d\tau, \quad (11a)$$

where $m-1 < \alpha < m$, and $\dfrac{\partial^\beta}{\partial|z|^\beta}$ is defined in Appendix B. In Eq. (11a), the relaxation effect depends on the local magnetization intensity at each instant of time. While, it may be possible that the relaxation does not depend on the local magnetization intensity and is independent of the diffusion induced attenuation, then the modified-Bloch equation could be written as

$$M(z,t) = \exp\left(-\frac{t}{T_2}\right)\left( \sum_{k=0}^{m-1} M^{(k)}(z,0^+)\frac{t^k}{k!} + \int_0^t \left\{ \left( \frac{(t-\tau)^{\alpha-1}}{\Gamma(\alpha)} D_{f_2} \frac{\partial^\beta}{\partial|z|^\beta} - i\gamma \mathbf{g}(t)\cdot z \right) M(z,t) \right\} d\tau \right). \quad (11b)$$

Whether or not the relaxation effect depends on the local magnetization intensity in anomalous diffusion is still not clear, which requires future experimental study to clarify. In the rest of this paper, only the relaxation effect described by Eq. (11a) will be considered. From Eq. (11a), the three-dimensional fractional integral modified Bloch equation can be written as

$$M(\mathbf{r},t) = \sum_{k=0}^{m-1} M^{(k)}(\mathbf{r},0^+)\frac{t^k}{k!} + \int_0^t \left\{ \left( \frac{(t-\tau)^{\alpha-1}}{\Gamma(\alpha)} D_{f_2} \nabla^\beta_{f_2} - i\gamma \mathbf{g}(t)\cdot\mathbf{r} - \frac{1}{T_2} \right) M(\mathbf{r},t) \right\} d\tau. \quad (12)$$

where $\nabla^\beta_{f_2} = \dfrac{\partial^\beta}{\partial|x|^\beta} + \dfrac{\partial^\beta}{\partial|y|^\beta} + \dfrac{\partial^\beta}{\partial|z|^\beta}$. For anisotropic diffusion, the integral modified-Bloch equation based on the fractional derivative in the principal axis frame will be

$$M'(\mathbf{r}',t) = \sum_{k=0}^{m-1} M'''^{(k)}(\mathbf{r}',0^+)\frac{t^k}{k!} +$$

$$\int_0^t \left\{ \left[ \frac{(t-\tau)^{\alpha-1}}{\Gamma(\alpha)} \left( D'_{f_2 x'x'}\frac{\partial^{\beta_{x'}}}{\partial|x'|^{\beta_{x'}}} + D'_{f_2 y'y'}\frac{\partial^{\beta_{y'}}}{\partial|y'|^{\beta_{y'}}} + D'_{f_2 z'z'}\frac{\partial^{\beta_{z'}}}{\partial|z'|^{\beta_{z'}}} \right) - i\gamma\mathbf{g}'(t)\cdot\mathbf{r} - \frac{1}{T_2} \right] M'(\mathbf{r}',t) \right\} d\tau.$$

(13)

For space-fractional diffusion, whose $\alpha = 1$, Eq. (13) reduces to

$$\frac{\partial}{\partial t}M'(\mathbf{r}',t) = \left[ \left( D'_{f_2 x'x'}\frac{\partial^{\beta_{x'}}}{\partial|x'|^{\beta_{x'}}} + D'_{f_2 y'y'}\frac{\partial^{\beta_{y'}}}{\partial|y'|^{\beta_{y'}}} + D'_{f_2 z'z'}\frac{\partial^{\beta_{z'}}}{\partial|z'|^{\beta_{z'}}} \right) - i\gamma\mathbf{g}'(t)\cdot\mathbf{r}' - \frac{1}{T_2} \right] M'(\mathbf{r}',t). \quad (14)$$

The solution of Eq. (14) can be assumed as $M'(\mathbf{r}',t) = S(t)\exp(-i\mathbf{K}'(t)\cdot\mathbf{r}')\exp(-\frac{t}{T_2})$, which can be substituted into Eq. (14) to give

$$S(t) = \exp(-\frac{t}{T_2})\exp\left( -\sum_{i'=x',y',z'} \int_0^t D'_{f_2 i'i'} K'^{\beta_{i'}}_{i'}(t)dt \right). \quad (15)$$

For PGSE or PGSTE experiments, Eq. (15) can be integrated to give

$$S(t) = \exp(-\frac{t}{T_2})\exp\left[ -\sum_{i=x',y',z'} D'_{f_2 ii}(\gamma g'_i\delta)^{\beta_i}(\Delta - \frac{\beta_i - 1}{\beta_i + 1}\delta) \right]. \quad (16)$$

For a system with homogeneous spin density, the magnetization of a general fractional diffusion in the principal axis frame can be described as $M'(\mathbf{r}',t) = S(t)\exp(-i\mathbf{K}'(t)\cdot\mathbf{r}')$. Substituting $M'(\mathbf{r}',t) = S(t)\exp(-i\mathbf{K}'(t)\cdot\mathbf{r}')$ into Eq. (13), we get

$$S(t) = \sum_{k=0}^{m-1} S^{(k)}(0)\frac{t^k}{k!} + \int_0^t \left\{ -\frac{(t-\tau)^{\alpha-1}}{\Gamma(\alpha)} \left[ D'_{f_2 x'x'}|k'_{x'}(t)|^{\beta_{x'}} + D'_{f_2 y'y'}|k'_{y'}(t)|^{\beta_{y'}} + D'_{f_2 z'z'}|k'_{z'}(t)|^{\beta_{z'}} \right] - \frac{1}{T_2} \right\} S(\tau)d\tau.$$

(17)

Neglecting the $T_2$ relaxation, we have

$$_t D^\alpha_* S(t) = -\left( D'_{f_2 x'x'}|k'_{x'}(t)|^{\beta_{x'}} + D'_{f_2 y'y'}|k'_{y'}(t)|^{\beta_{y'}} + D'_{f_2 z'z'}|k'_{z'}(t)|^{\beta_{z'}} \right) S(t). \quad (18)$$

The similar type of fractional equation to Eq. (18) has been solved by the Adomian decomposition method [51-55]. According to these references, the solution of Eq. (18) is [46,51-55]

$$S(t) = \sum_{n=0}^{\infty} S_n(t), \quad (19a)$$

where

$$S_0(t) = \sum_{k=0}^{m-1} S^{(k)}(0^+)\frac{t^k}{k!}, \quad m-1 < \alpha < m, \quad (19b)$$

and

$$S_n(t) = J^\alpha\left(a(t)S_{n-1}(t)\right)$$

$$= -\int_0^t \frac{(t-\tau)^{\alpha-1}}{\Gamma(\alpha)}\left[\sum_{i=x',y',z'} D'_{f_2 ii}K_i^{\prime\beta_i}(t')\right]S_{n-1}(\tau)d\tau,$$

$$= -\int_0^t \frac{\left[\sum_{i=x',y',z'} D'_{f_2 ii}K_i^{\prime\beta_i}(t')\right]S_{n-1}(\tau)d(t-\tau)^\alpha}{\alpha\Gamma(\alpha)}$$

(19c)

where $a(t) = -\sum_{i=x',y',z'} D'_{f_2 ii}K_i^{\prime\beta_i}(t')$. When $S^{(k)}(0^+) = 0$ [15,17,46], and $S_0(t) = 1$ are used, because in typical PGSE or PGSTE experiments $K'_i(\tau) = K'_i(t_{tot} - \tau)$, at small attenuation, we have

$$S(t) \approx S_0(t) + S_1(t)$$

$$= 1 - \int_0^t \frac{(t-\tau)^{\alpha-1}}{\Gamma(\alpha)}\left[\sum_{i=x',y',z'} D'_{f_2 ii}K_i^{\prime\beta_i}(t')\right]d\tau$$

$$\approx E_{\alpha,1}\left(-\sum_{i=x',y',z'}\int_0^t D'_{f_2 ii}K_i^{\prime\beta_i}(t')dt'^\alpha\right)$$

$$\approx E_{\alpha,1}(-\mathbf{b}'_f : \mathbf{D}'_{\mathbf{f}_2})$$

(20)

where $\mathbf{b}'_f$ is defined by Eq. (10a). Thus, the results from the modified Bloch equation and that from ISA method approximation are close to each other at a small level of signal attenuation.

2.1.3 Complicated coupled case

It is still possible that the diffusions in three dimensions have different $D_{f_{i'}}$, $\alpha_{i'}$, $\beta_{i'}$ but are coupled in a certain manner. For instance, a correlated random walk takes place in three directions. The possibility of a given jump having a direction along one of three principal axis is proportional to its total time staying in that axes in the whole diffusion process. The total diffusion time $t = \sum_{i'=1} t_{i'}$, $i' = x', y', z'$ where $t_{i'}$ is the total time on $i'$ axis. For the fractal derivative model, by setting $c_{i'} = t_{i'}/t$ (or $c_{i'} = 3t_{i'}/t$, which is consistent with the simple case in section 2.1.1), $dt_{i'}^{\alpha_{i'}} = \alpha_{i'}c_{i'}^{\alpha_{i'}}t^{\alpha_{i'}-1}dt$, the attenuation may be obtained based on Eq. (9) as

$$S(t) = \exp(-\frac{t}{T_2})\exp\left[-\sum_{i'=x',y',z'}\int_0^t \alpha_{i'}c_{i'}^{\alpha_{i'}}t^{\alpha_{i'}-1}D'_{f_i i'i'}K_{i'}^{\prime\beta_{i'}}(t)dt\right].$$

(21)

For the fractional derivative model, by setting $c_{i'} = t_{i'}/t$ (or $c_{i'} = 3t_{i'}/t$), we have $(t_{i'} - \tau_{i'})^{\alpha_{i'}-1}d\tau_{i'} = c_{i'}[c_{i'}(t-\tau)]^{\alpha_{i'}-1}d\tau$. Thus, based on Eq. (19), the signal attenuation may be given as

$$S(t) = \sum_{n=0}^{\infty} S_n(t), \tag{22a}$$

where

$$S_0(t) = \sum_{k=0}^{m-1} S^{(k)}(0^+)\frac{t^k}{k!}, m-1 < \alpha < m, \tag{22b}$$

and

$$S_n(t) = J^\alpha\left(a(t)S_{n-1}(t)\right) = -\int_0^t \left[\sum_{i'=x',y',z'} \frac{c_{i'}[c_{i'}(t-\tau)]^{\alpha_{i'}-1}d\tau}{\Gamma(\alpha_{i'})} D'_{f_2 i'i'} K'^{\beta_{i'}}_{i'}(t')\right] S_{n-1}(\tau)d\tau. \tag{22c}$$

2.2 Uncoupled anisotropic diffusion with different $D_{f_{i'}}$, $\alpha_{i'}$, $\beta_{i'}$ on $x', y', z'$ axes

For the uncoupled diffusion, the diffusions in $x', y', z'$ axes are independent of each other and all the three anomalous diffusion parameters, $D_{f_{i'}}$, $\alpha_{i'}$, $\beta_{i'}$ on the three axes in the principal frame may be different. In such a situation, the signal attenuation can be obtained by

$$S(t) = S_{x'}(t)S_{y'}(t)S_{z'}(t), \tag{23}$$

where $S_{i'}(t)$ is the attenuation expression of one-dimensional anomalous diffusion, which can be reduced from Eqs. (9) and (19). For the fractal derivative model, from Eq. (9), we have

$$S_{i'}(t) = \exp\left[-\int_0^t \alpha_{i'} t^{\alpha_{i'}-1} D'_{fi'i'} K'^{\beta_{i'}}_{i'}(t)dt\right], \tag{24}$$

while for the fractional derivative model, from Eq. (19), we get

$$S_{i'}(t) = \sum_{n=0}^{\infty} S_n(t), S_0(t) = \sum_{k=0}^{m-1} S^{(k)}(0^+)\frac{t^k}{k!}, m-1 < \alpha < m, S_n(t) = -\int_0^t \frac{D'_{f_2 i'i'} K'^{\beta_{i'}}_{i'}(t') S_{n-1}(\tau) d(t-\tau)^{\alpha_{i'}}}{\alpha_{i'}\Gamma(\alpha_{i'})}. \tag{25}$$

As $\exp(-x-y-z) = \exp(-x)\exp(-y)\exp(-z)$, the stretched exponential attenuation of such uncoupled anisotropic diffusion will obey the same equation as Eq. (9). While, for the diffusion described based on the fractional derivative model, as $E_{\alpha,1}(-x-y-z) \neq E_{\alpha,1}(-x)E_{\alpha,1}(-y)E_{\alpha,1}(-z)$, the signal attenuation based on Eqs. (23) and (25) is different from that based on Eqs. (19a-c).

3. CTRW simulation

The continuous time random walk has been employed to simulate PFG isotropic fractional diffusion [33,36]. The simulation will be modified to verify the theoretical results for anisotropic fractional diffusion. In the simulation, the random walk of a spin in the real space consists of a sequence of independent random waiting times $\Delta t_1$, $\Delta t_2$, $\Delta t_3$, ..., $\Delta t_n$, and a sequence of random jumps $\Delta \xi_{i'1}$, $\Delta \xi_{i'2}$, $\Delta \xi_{i'3}$, ..., $\Delta \xi_{i'n}$, where $\xi_{i'} = x', y', z'$. Two models, CTRW model [47] and Lattice model [48,49], were used in the simulation.

The CTRW model described in reference [47] is used to produce the waiting time $\tau$ and jump length $\varepsilon$ for each random jump, which is proposed to model different fractional diffusion systems in physics and economics [47]. The waiting time distribution in the simulation follows a Mittag-Leffler function which can be approximated as a stretched exponential function at small $t$ value, and a power law at large $t$ value [47]. The waiting time $\tau$ is given by [47]

$$\tau = -\eta_t \log U \left( \frac{\sin(\alpha_{i'}\pi)}{\tan(\alpha_{i'}\pi V)} - \cos(\alpha\pi) \right)^{\frac{1}{\alpha_{i'}}}, \tag{26}$$

where $\eta_t$ is a scale constant, and $U, V \in (0,1)$ are two independent random numbers. While the jump length distribution behaves as a Levy $\alpha$-stable function which is a generalization of a Gaussian function [47]. The jump length $\varepsilon_{i'}$ is given by [47]

$$\varepsilon_{i'} = \eta_{i'} \left( \frac{-\log U \cos(\Phi)}{\cos((1-\beta_{i'})\Phi)} \right)^{1-\frac{1}{\beta_{i'}}} \frac{\sin(\beta_{i'}\Phi)}{\cos(\Phi)}, \tag{27}$$

where $i' = x', y', z'$, $\Phi = \pi(V-1/2)$ and $\eta_{i'}$ is a scale constant. Reference [47] shows that the probability density function based on the CTRW model satisfies the fractional diffusion equation in the diffusive limit. In the simulation, for the coupled anisotropic diffusion, there are equal possibilities in the $x', y', z'$ directions and each jump will take one of those directions (the complicated anisotropic anomalous diffusion in section 2.1.3 is not simulated here). While, for the uncoupled or non-correlated anisotropic diffusion, three independent random walks in $x', y', z'$ directions take place simultaneously.

Additionally, the algorithm used in the lattice model simulation from references [48,49] was modified to record the time $t_j = \sum_{k=1}^{j} \tau_k$ and displacement $\xi_{i'}(t_j) = \sum_{k=1}^{j} \varepsilon_{i'k}$ of the CTRW random walk. The net change of spin phase $\phi_m(t)$ in the $m^{th}$ walk is [33,36,48-50]

$$\phi_m(t) = \sum_{i'=x',y',z'} \sum_{j=1}^{n} \gamma g'_{i'}(t_j) \xi_{i'}(t_j) \tau_j, \tag{28}$$

where $g'_{i'}(t_j)$ is the gradient strength at the time $t_j$, and $j$ is the time index from 1 to $n$. In the simulation, a discrete time $t'_{j'} = \sum_{k=1}^{j'} \Delta t$ with equal time interval $\Delta t$ was used to record the continuously evolving phase, which is calculated by $\phi_m(t'_{j'}) = \sum_{i'=x',y',z'} \left[ \sum_{j=1}^{l} \gamma g'_{i'}(t_j) \xi_{i'}(t_j) \tau_j + \gamma g'_{i'}(t_l) \xi_{i'}(t_l)(t'_{j'} - t_l) \right]$ with $t_l \leq t'_{j'} \leq t_{l+1}$. The term $\gamma g'_{i'}(t_l) \xi_{i'}(t_l)(t'_{j'} - t_l)$ represents the partial phase evolution of the $l+1^{th}$ jump step that the recorded time $t'_{j'}$ belongs. The manner of phase recording does not affect the phase evolution process. Thus, either a discrete recording or a continuous time recording can be used. The total normalized signal attenuation can be directly obtained from the simulation via [49,50]

$$S(t) = \frac{1}{N_{walks}} \sum_{m=1}^{N_{walks}} \cos[\phi_m(t)]. \quad (29)$$

A total of at least 100,000 walks were performed for each simulation. Only the subdiffusion based on the fractional derivative model was simulated because the model in reference [47] is proposed specifically for subdiffusion based on the fractional derivative.

## 4. Results and Discussion

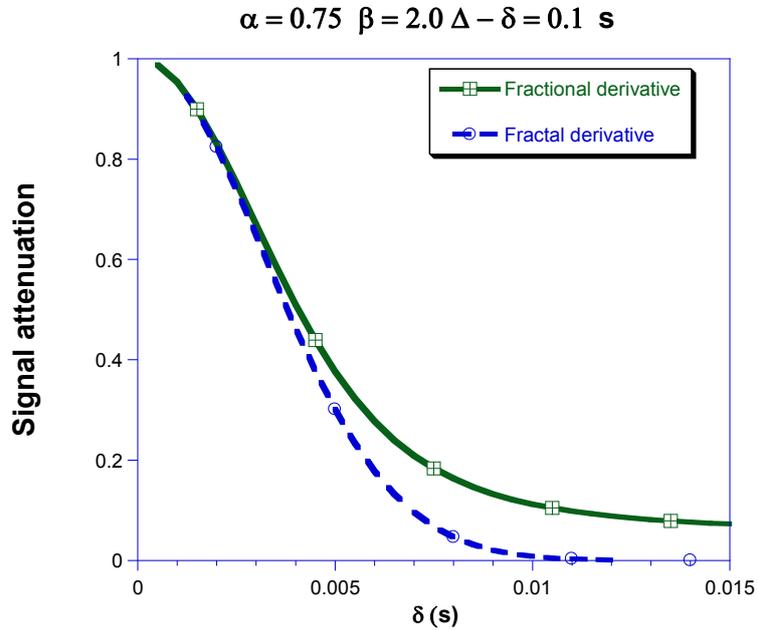

**Fig. 2** Comparison of the PFG signal attenuation from the modified-Bloch equations based on the fractal derivative Eq. (9) with these based on the fractional derivative Eq. (19). The parameters used in these theoretical predictions are $\alpha_{i'} = 0.75, \beta_{i'} = 2$, $D'_{f_2 x'x'} + D'_{f_2 y'y'} + D'_{f_2 z'z'} = 1.426 \times 10^{-9}$ m$^\beta$/s$^\alpha$, $D'_{f_2 t't'} = D'_{f_1 t't'}/\Gamma(1+\alpha)$, and $g'_{i'} = 0.05$ T/m. Relaxation effect is neglected.

The recently developed modified-Bloch equation method [46] was extended to give the general PFG signal attenuation expressions including the FGPW effect for free anisotropic anomalous diffusion. The modified-Bloch equations are based on two different models, the fractal derivative model and the fractional derivative model. The attenuation based on the fractal derivative model and that based on the fractional derivative model have similarities and differences. The fractal differential modified-Bloch equation yields a stretched exponential function (SEF) attenuation, while the fractional integral modified-Bloch equation yields a Mittag-Leffler function (MLF) based signal attenuation. The similarity and difference between SEF and MLF attenuations are shown in Fig. 2. At small level of signal attenuation, the MLF can be approximated as a SEF, while at large signal attenuation, the MLF attenuates more slowly than the SEF at subdiffusion. Only the subdiffusion is shown in Fig. 2, the superdiffusion is more complicated, and therefore not the focus of this paper.

The results of space-fractional diffusion agree with those obtained by the different modified Bloch equations proposed by other groups [37-39]. For general fractional diffusion, the results based on the fractional integral modified-Bloch equation agree with the previously reported result based on a different modified-Bloch equation in Ref. [37] at the first gradient pulse (The reported result in Ref. [37] only gives the attenuation in the first gradient pulse). The results can be reduced to the reported anisotropic normal diffusion result when $\alpha =1$, and $\beta = 2$. These results can also be reduced to the reported isotropic anomalous diffusion result when $\alpha_{x'} = \alpha_{y'} = \alpha_{z'} = \alpha$ and $\beta_{x'} = \beta_{y'} = \beta_{z'} = \beta$. Both the results can be reduced to the one-dimensional PFG anomalous diffusion results based on the modified-Bloch equation in Ref. [46]. The results from the modified-Bloch equation method also agree with the effective phase shift diffusion method and the observing the signal intensity at the origin method (see Appendix D).

The results from the modified-Bloch equation are in good agreement with the CTRW simulations. Fig. 3 shows the comparison between the theoretical predicting based on Eq. (19) with the CTRW simulation for coupled anisotropic diffusion with same derivative parameters but different diffusion constant components, where $\alpha = 0.75$, $\beta = 2$ are used for all three axes, and $D'_{f_2 x'x'} : D'_{f_2 y'y'} : D_{f_2 z'z'} = 1:4:0.25$ with $D'_{f_2 x'x'} = 2.72 \times 10^{-10}$ m$^2$/s$^{0.75}$. Fig. 4 shows the comparison of the theoretical prediction based on Eq. (19) with CTRW simulation for coupled anisotropic diffusion with the same $\alpha_{i'}$ but different $\beta_{i'}$ and diffusion constant components, where $\alpha_{i'} = \alpha = 0.75$, $\beta_{x'} = \beta_{y'} = 1.5$, $\beta_{z'} = 1.9$, $D'_{f_2 x'x'} = D'_{f_2 y'y'} = 4.51 \times 10^{-7}$ m$^{1.5}$/s$^{0.75}$, and $D_{f_2 z'z'} = 1.10 \times 10^{-9}$ $m^{1.9}$/s$^{0.75}$.

Fig. 3(a)

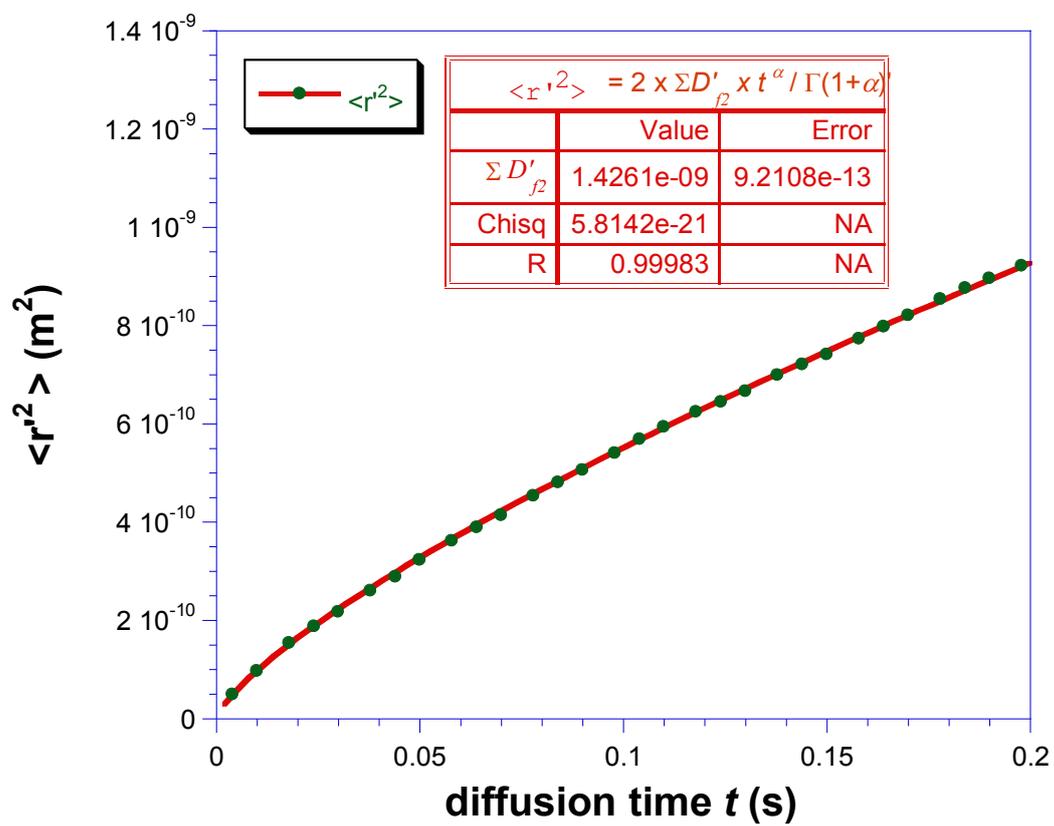

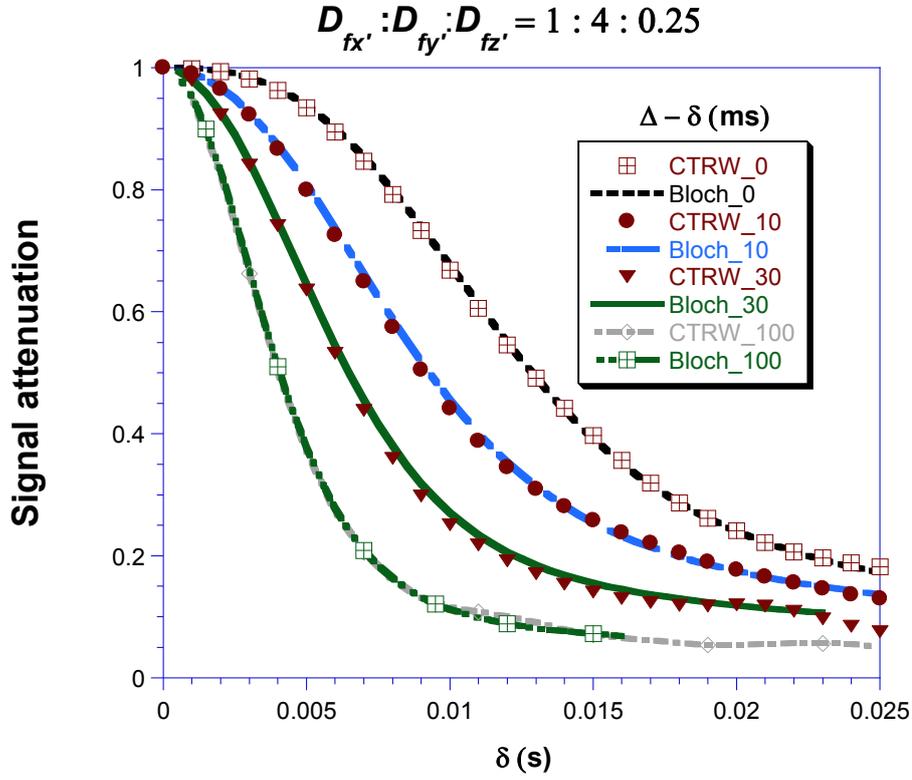

Fig. 3 Comparison of the PFG signal attenuation predicted from the modified-Bloch equation with the CTRW simulation results for anomalous diffusion with the same variables $(\alpha_{i'}, \beta_{i'})$ but different $D'_{f_{2i'i'}}$ ( $\alpha_{i'} = 0.75, \beta_{i'} = 2$, and $D'_{f_{2x'x'}} : D'_{f_{2y'y'}} : D'_{f_{2z'z'}} = 1:4:0.25$ ): (a) $\langle r'^2(t) \rangle$ versus $t$ from the simulation, the fitting gives $\sum D'_{f_2} = D'_{f_{2x'x'}} + D'_{f_{2y'y'}} + D'_{f_{2z'z'}} = 1.426 \times 10^{-9}$ m$^\beta$/s$^\alpha$, (b) finite gradient pulse width effect with $\Delta - \delta$ equaling 0 ms, 10 ms, 30 ms and 100 ms, $g'_{i'}$ equaling 0.05 T/m.

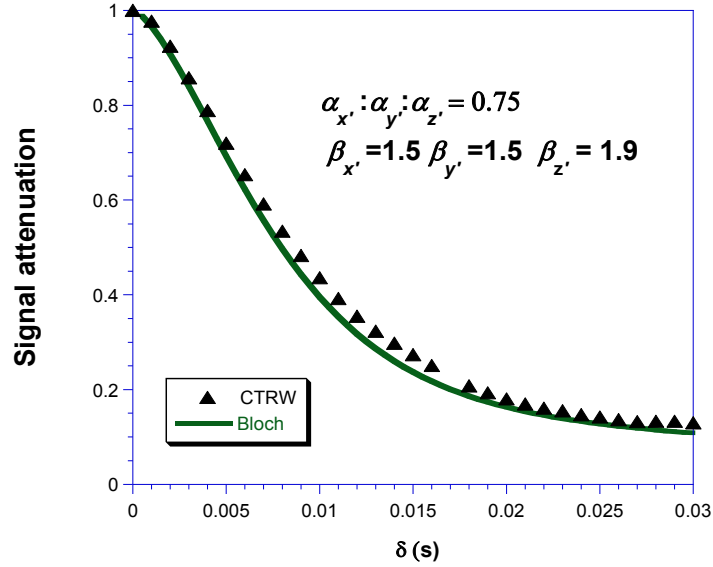

**Fig. 4** Comparison of the PFG signal attenuation predicted from the modified-Bloch equation with the CTRW simulation results for anisotropic anomalous diffusion with same $\alpha_{i'}$ but different $\beta_{i'}$ and $D'_{f_2 i' i'}$. The parameters are $\alpha_{i'} = 0.75$, $\beta_{x'} = \beta_{y'} = 1.5$, $\beta_{z'} = 1.9$, $D'_{f_i x' x'} = D'_{f_i y' y'} = 4.51 \times 10^{-7}$ m$^{1.5}$/s$^{0.75}$, $D'_{f_i z' z'} = 1.10 \times 10^{-9}$ m$^{1.9}$/s$^{0.75}$, $\Delta - \delta = 100$ ms and $g'_{i'}$ equaling 0.0125 T/m.

There is a clear difference between the signal attenuation of coupled and uncoupled fractional diffusion based on the fractional derivative model. Fig. 5 shows the comparison of CTRW simulation of uncoupled diffusion with the theoretical predictions based on Eqs. (23) and (25) for uncoupled diffusion and Eq. (19) for coupled diffusion. From Fig. 5, the prediction from the coupled attenuation Eq. (19) is smaller than that of the CTRW simulation at large attenuation, while the signal attenuation from the Eqs. (23) and (25) for the uncoupled anisotropic diffusion agrees well with the CTRW simulation across the whole curve. The difference between the coupled and uncoupled anomalous diffusion should arise from that

$$E_{\alpha,1}(-\sum_{i'=x',y',z'}\int_0^t D'_{f_2 i' i'} K'^{\beta_{i'}}_{i'}(t')dt'^{\alpha_{i'}}) \neq \prod_{i'=x',y',z'} E_{\alpha,1}(-\int_0^t D'_{f_2 i' i'} K'^{\beta_{i'}}_{i'}(t')dt'^{\alpha_{i'}})$$

, which is significantly different from the SEF attenuation. Fig. 5 indicates that in real applications, it may need to tell coupled or uncoupled diffusions.

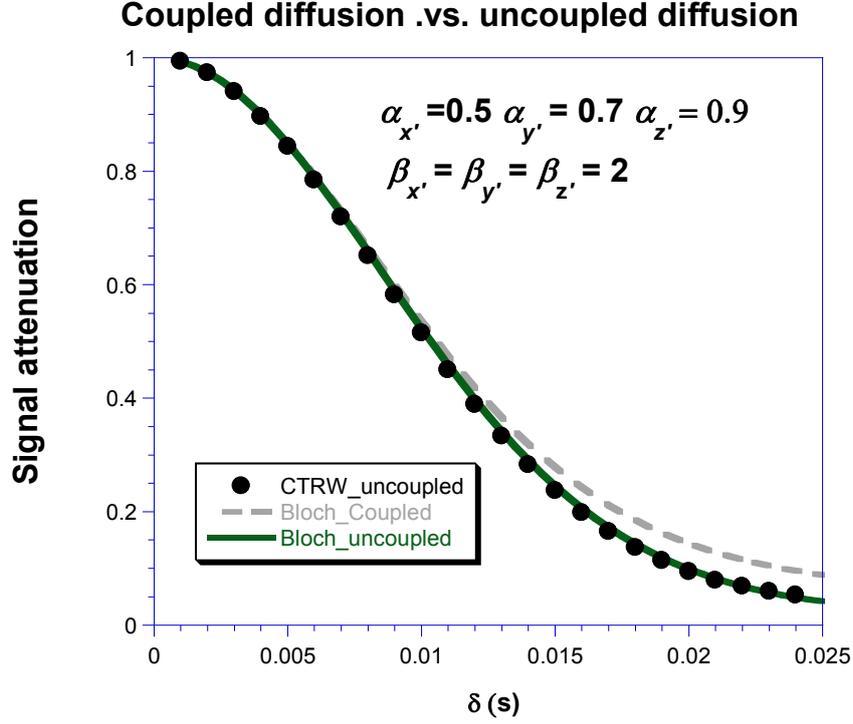

**Fig. 5** Comparison the PFG signal attenuation between uncoupled and coupled anisotropic anomalous diffusion from CTRW and theoretical predictions. The theoretical predictions for coupled diffusion are calculated by Eq. (19), while the theoretical prediction for uncoupled diffusion is calculated by Eqs. (23) and (25). The parameters used are $\alpha_{x'} = 0.5$, $\alpha_{y'} = 0.7$, $\alpha_{z'} = 0.9$, $\beta_{x'} = \beta_{y'} = \beta_{z'} = 2$, $D'_{f_2 x'x'} = 6.47 \times 10^{-11}$ m$^2$/s$^{0.5}$, $D'_{f_2 y'y'} = 4.92 \times 10^{-10}$ m$^2$/s$^{0.7}$, $D'_{f_2 z'z'} = 3.59 \times 10^{-9}$ m$^2$/s$^{0.9}$, $\Delta - \delta = 100$ ms and $g'_{i'}$ equaling 0.0125 T/m.

It is worth noting that $\mathbf{D_f} = \widetilde{R} \cdot \mathbf{D'_f} \cdot R$ may only be used under the condition $\alpha_{x'} = \alpha_{y'} = \alpha_{z'} = \alpha$, $\beta_{x'} = \beta_{y'} = \beta_{z'} = \beta$. If this condition does not hold, the units of $D'_{f i'i'}$, $i' = x', y', z'$ are $m^{\beta_{i'}}/s^{\alpha_{i'}}$ which are different among the three principal axes. Therefore, one may need to transfer $\mathbf{g}$ to $\mathbf{g'} = R\mathbf{g}$ and calculate the PFG signal attenuation in the principal axis frame directly. This is different from anisotropic normal diffusion.

In real diffusion systems, the parameters $\alpha_{i'}$ and $\beta_{i'}$ are related with the waiting time distribution and jump length distribution, which are affected by the material morphology and dynamic property. The parameters $\alpha_{i'}$ and $\beta_{i'}$ could be experimentally determined in PFG experiments. At small level of signal attenuation, the signal attenuation expression Eqs. (19) and (20) is equivalent to Eq. (9), namely,

$$E_{\alpha,1}(-\mathbf{b}'_f : \mathbf{D'_{f_2}}) \approx \exp\left[-\left(\mathbf{b}'_f : \mathbf{D'_{f_2}}\right)/\Gamma(1+\alpha)\right], \tag{30}$$

when $\left|\mathbf{b}'_f : \mathbf{D}'_{f_2}\right|$ is small. Based on Eq. (30), at small signal attenuation, we have

$$\ln[\ln(A(0) - \ln A(t)] = \begin{cases} c_1 + \beta_i \ln(g'_i), & \text{when } \delta, \Delta \text{ and } g'_{j \neq i} \text{ are fixed} \\ c_2 + (\alpha + \sum_i \varepsilon_i \beta_i) \ln(\delta), & \text{when } \Delta = \delta, \text{ and } g' \text{ is fixed} \\ c_3 + \alpha \ln(\Delta), & \text{when } \delta \ll \Delta, \text{ and } g' \text{ is fixed} \end{cases}, \quad (31)$$

where $c_i, i = 1,2,3$ is a constant, and $\varepsilon_i = \begin{cases} 1, g'_i \neq 0 \\ 0, g'_i = 0 \end{cases}$. When the linear fitting is used to fit experimental data based on Eq. (31), the signal attenuation should be small. However, the small signal attenuation requirement may be slightly relaxed by performing a polynormal fitting of the curve, which can give the $\alpha_{i'}$ and $\beta_{i'}$ values from the coefficient of the first order term. From Ref. [33], the determined $\beta_{i'}$ values based on simulation or experimental data are within a narrow range, while the $\alpha_{i'}$ values show a large difference between the fittings based on the SEF and MLF attenuations (0.62 and 0.7 for MLF, while 0.81 for SEF), which is due to the fact that SEF attenuation is faster than MLF attenuation [18]. Because the $\beta_{i'}$ values could be determined within a narrow range, it should be determined from the small signal attenuation experimental data first, which may not only be easier but could also improve the accuracy of analysis. Then, the obtained $\beta_{i'}$ values can be used to determine the $\alpha_{i'}$ values from the large signal attenuation experimental data. Additionally, $\alpha_{i'}$ values can be determined by other methods from experimental data. For example, in reference [36], it is found that using $z^\beta$ vs. $\sqrt{D_f t^\alpha}$ to fitting the restricted diffusion data can yield a $\alpha$ value within a narrow range. Generally, in real applications, the parameters $\alpha_{i'}$ and $\beta_{i'}$ should not be arbitrarily selected.

The modified-Bloch equation is very similar to the fractional reaction-diffusion equation [56]. These two terms $-i\gamma(\mathbf{g}(t) \cdot \mathbf{r})M(\mathbf{r},t)$ and $\frac{M(\mathbf{r},t)}{T_2}$ may be viewed as "reaction" terms. In a spin system, the gradient field $\mathbf{g}(t)$ and the spin-spin relaxation $T_2$ affect the phase evolution of spin moments, but these two "reactions" always take place in the NMR sample without regard to how fast or slow the translational diffusion is. In fact, the phase evolution affected by the gradient field can be viewed as a phase diffusion process which can be explained by the effective phase shift diffusion equation method [18]. The modified-Bloch equation is a macroscopic approach that includes both the translational diffusion and the phase evolution of a spin system.

The numerical evaluation of the PFG signal attenuation can be performed with the computer assistance. The calculation of the stretched exponential function based PFG signal attenuation is easier than that of Mittag-Leffler function based PFG signal attenuation. Besides the Adomian decomposition method [51-55], the direct integration method (see Appendix C) can be used to numerically evaluate the Mittag-Leffler function based PFG signal attenuation. The results from the direct integration method agree with those obtained by the Adomian decomposition method as shown in Fig. 6a. The direct integration method is much faster than the Adomian decomposition method. Additionally, the direct integration method can be used to calculate Mittag-Leffler type functions $E_{\alpha,1}(-ct^\alpha)$ and $E_{\alpha,1}(-t^\alpha)$. The Mittag-Leffler function calculated by Eq (C.3) agree with those obtained by the method in Ref. [57] and the Pade approximation [58], which is shown in Fig. 6b. The FORTRAN code for the calculation based on Eq. (C.3) and Pade approximation can be obtained from Ref. [59]. This direct integration calculation method is simple and fast. Especially, it does not cause overflow in computation.

**Fig. 6a**

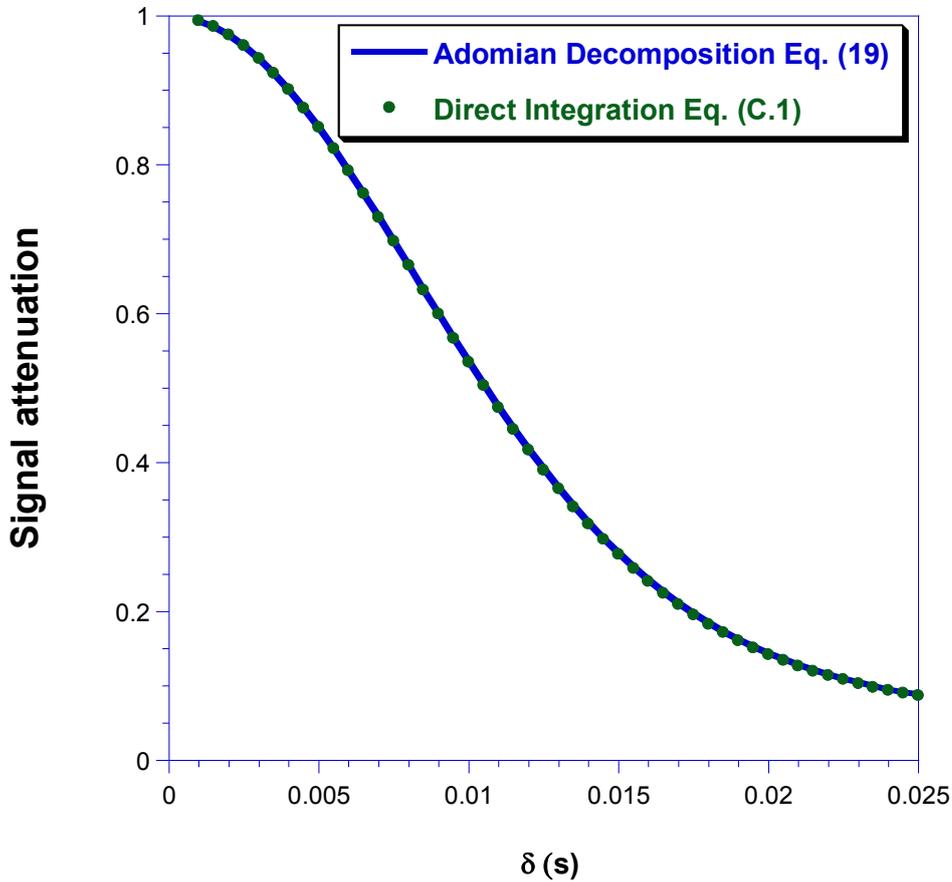

**Fig. 6b**

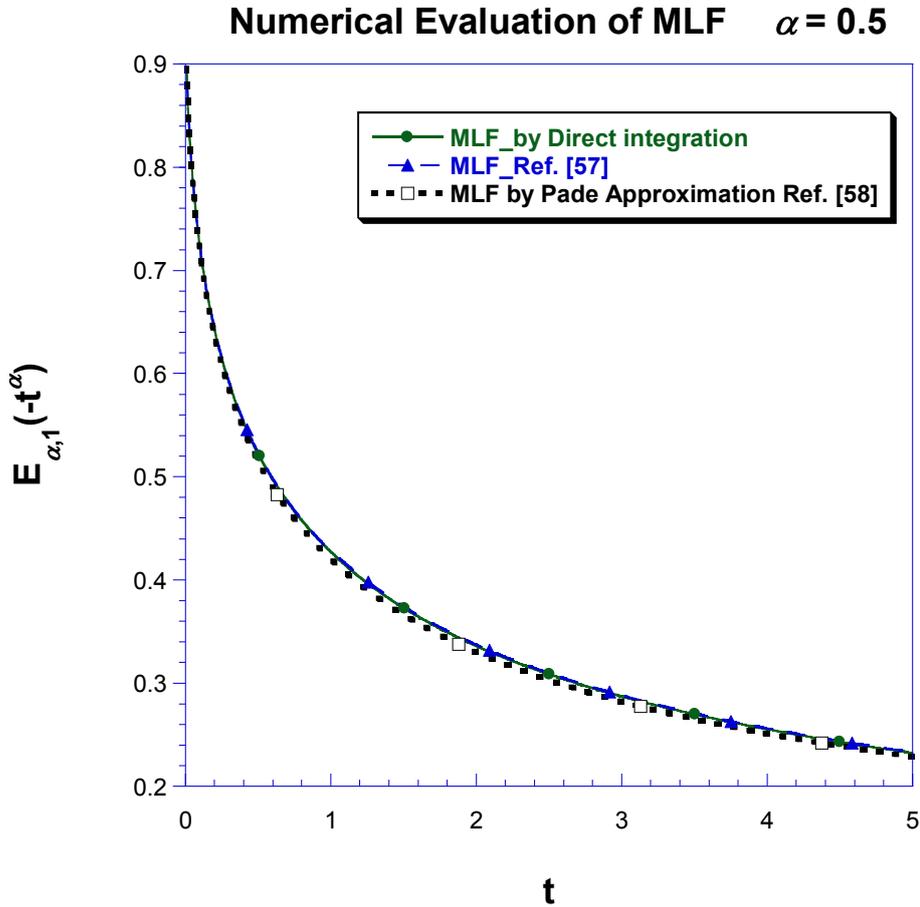

Fig. 6 The numerical evaluation by the direct integration method (DIM): (a) the good agreement between the direct integration method and the Adomian decomposition method in the numerical evaluation of PFG signal attenuation, the parameters used are $\alpha_{x'} = 0.5$, $\alpha_{y'} = 0.7$, $\alpha_{z'} = 0.9$, $\beta_{x'} = \beta_{y'} = \beta_{z'} = 2$, $D'_{f_2 x'x'} = 6.47 \times 10^{-11}$ m²/s$^{0.5}$, $D'_{f_2 y'y'} = 4.92 \times 10^{-10}$ m²/s$^{0.7}$, $D'_{f_2 z'z'} = 3.59 \times 10^{-9}$ m²/s$^{0.9}$, $\Delta - \delta = 100$ ms and $g'_{i'}$ equaling 0.0125 T/m, (b) the good agreement of the direct integration method in calculation of the Mittag-Leffler function $E_{\alpha,1}(-t^\alpha)$ between the direct integration method with these by method in Ref. [57] and Pade approximation method in Ref. [58].

The PFG anisotropic anomalous diffusion results provide new formalisms for PFG anomalous diffusion studies in NMR and MRI. In particular, it can potentially be a convenient tool in diffusion tensor imaging (DTI) [60]. In real application systems, the diffusion and relaxation mechanism are complicated, which requires much effort to advance the theoretical and experimental studies in this field.

**Appendix A. Definition of fractal derivative from reference [11-12].**

$$\frac{\partial P^{\beta/2}}{\partial t^{\alpha}} = \lim_{t_1 \to t} \frac{P^{\beta/2}(t_1) - P^{\beta/2}(t)}{t_1^{\alpha} - t^{\alpha}}, 0 < \alpha, 0 < \beta/2. \tag{A.1}$$

**Appendix B. Definition of fractional derivative [3,15-17].**

The Caputo fractional derivative $_t D_*^{\alpha}$ is defined as [15-17]

$$_t D_*^{\alpha} f(t) := \begin{cases} \dfrac{1}{\Gamma(m-\alpha)} \displaystyle\int_0^t \dfrac{f^{(m)}(\tau) d\tau}{(t-\tau)^{\alpha+1-m}}, m-1 < \alpha < m, \\ \dfrac{d^m}{dt^m} f(t), \alpha = m. \end{cases} \tag{B.1}$$

The space fractional derivative is defined as [3,15-17]

$$\frac{d^{\beta}}{d|z|^{\beta}} = -\frac{1}{2\cos\frac{\pi\alpha}{2}} \left[ _{-\infty}D_z^{\beta} + _z D_{\infty}^{\beta} \right], \tag{B.2}$$

where

$$_{-\infty}D_z^{\beta} f(z) = \frac{1}{\Gamma(m-\beta)} \frac{d^m}{dz^m} \int_{-\infty}^{z} \frac{f(y)dy}{(z-y)^{\beta+1-m}}, \beta > 0, m-1 < \beta < m, \tag{B.3}$$

and

$$_z D_{\infty}^{\beta} f(z) = \frac{(-1)^m}{\Gamma(m-\beta)} \frac{d^m}{dz^m} \int_z^{\infty} \frac{f(y)dy}{(y-z)^{\beta+1-m}}, \beta > 0, m-1 < \beta < m. \tag{B.4}$$

**Appendix C. Direct integration method for numerical evaluation of Mittag-Leffler type PFG signal attenuation.**

The Adomian decomposition method provides an analytical approximate solution to Eq. (17). However, a direct integration method can be proposed as an alternate way for the numerical calculation of PFG signal attenuation. The algorithm for the direct integration is simple: first, divide the time into $t_j = \sum_{k=1}^{j} \Delta t_k$; second, calculate the

$$S(t_j) = 1 + \sum_{k=1}^{j} a(t_k) S(t_{k-1}) \left[ (t_j - t_{k-1})^{\alpha} - (t_j - t_k)^{\alpha} \right] / \Gamma(1+\alpha), a(t_k) < 0, \tag{C.1}$$

step by step from $j = 1$ to $n$, where $S(t_j)$ is the signal intensity at time $t_j$, and $\frac{(t-\tau)^{\alpha-1}}{\Gamma(\alpha)} dt = \frac{d(t-\tau)^{\alpha}}{\Gamma(1+\alpha)}$ is used in the calculation. In this calculation, each $S(t_j)$ is calculated only once, while in the Adomian decomposition method, $S(t) = \sum_{n=0}^{\infty} S_n(t), t = t_j$, which is a superposition of many terms. Thus, the calculation speed of the direct integration method can be a few orders of magnitude faster than the Adomian decomposition method. The results from the direct integration method agree with those obtained by the Adomian decomposition method as shown in Fig. 6a.

Moreover, under SGP approximation [18,24], $K'_i(t) = \gamma g'_i \delta$, $a(t) = -D'_{f_2 ii}(\gamma g'_i \delta)^{\beta_i} = -c$ is a constant; from Eqs. (17) and (19), we can get $S(t) = E_{\alpha,1}(-ct^\alpha)$, and $S(\tau) = E_{\alpha,1}(-c\tau^\alpha)$, which implies

$$E_{\alpha,1}(-ct^\alpha) = 1 - \int_0^t \left\{ \left( \frac{(t-\tau)^{\alpha-1}}{\Gamma(\alpha)} \right) cE_{\alpha,1}(-c\tau^\alpha) \right\} d\tau. \tag{C.2}$$

When $c = 1$, $E_{\alpha,1}(-ct^\alpha) = E_{\alpha,1}(-t^\alpha)$, therefore, the direct integration method can also be used to calculate Mittag-Leffler function and its derivative by the following discrete form of equations:

$$E_{\alpha,1}(-t_j^\alpha) = 1 + \sum_{k=1}^{j} a(t_k) E_{\alpha,1}(-t_{k-1}^\alpha) \left[ (t_j - t_{k-1})^\alpha - (t_j - t_k)^\alpha \right] / \Gamma(1+\alpha), a(t_k) < 0, \tag{C.3}$$

$$E'_{\alpha,1}(-t_j^\alpha) = \frac{\left[ E_{\alpha,1}(-t_j^\alpha) - E_{\alpha,1}(-t_{j-1}^\alpha) \right]}{\Delta t_j}. \tag{C.4}$$

The Mittag-Leffler function calculated by Eq. (C.3) agree with those obtained by the method in Ref. [57] and the Pade approximation [58], which is shown in Fig. 6b. The FORTRAN code for the calculation based on Eq. (C.3) and Pade approximation can be obtained from Ref. [58]. This direct integration calculation method is simple and fast. Especially, it has no overflow problem in computation.

**Appendix D. General PFG solutions by other methods:**

I.   EPSDE method:

For the fractal derivative model, the FGPW effect can be obtained by the effective phase shift diffusion equation method. The anisotropic effective phase shift diffusion equation can be written as [46]

$$\frac{\partial P'(\phi_{x'}, \phi_{y'}, \phi_{z'}, t)}{\partial t^\alpha} = \sum_{i'=x',y',z'} D'_{f_1 i'i'} K'^{\beta_{i'}}_{i'}(t) \frac{\partial}{\partial \phi_{i'i'}^{\beta_{i'}/2}} \left( \frac{\partial P'(\phi_{x'}, \phi_{y'}, \phi_{z'}, t)}{\partial \phi_{i'i'}^{\beta_{i'}/2}} \right), \tag{D.1}$$

where $D'_{f_1 i'i'} K'^{\beta_{i'}}_{i'}(t)$ is the effective phase diffusion coefficient. The solution of Eq. (D.1) can be obtained similarly to that of the one dimension diffusion in ref. [46] as

$$P'(\phi_{x'}, \phi_{y'}, \phi_{z'}, t) = \frac{c}{(4\pi)^{3/2} \sqrt{\prod_{i'=x',y',z'} \int_0^t D'_{f_1 i'i'} K'^{\beta_{i'}}_{i'}(t) dt^\alpha}} \exp\left[ -\sum_{i'=x',y',z'} \frac{\phi_{i'}^{\beta_i}}{4 \int_0^t D'_{f_1 i'i'} K'^{\beta_{i'}}_{i'}(t) dt^\alpha} \right]. \tag{D.2}$$

The signal attenuation can be obtained as [18,46]

$$S_{SGP}(t) = \int_{-\infty}^{\infty} P(\phi, t) \exp(+i\phi) d\phi = \exp\left[ -\sum_{i'=x',y',z'} \int_0^t D'_{f_1 i'i'} K'^{\beta_{i'}}_{i'}(t) dt^\alpha \right], \tag{D.3}$$

which can be reduced to the one-dimensional anomalous diffusion result in Refs. [18,33,46] and the anisotropic normal diffusion result when $\alpha = 1, \beta = 2$ in Ref. [24,57].

While, for the fractional derivative model, the phase diffusion equation based on the fractional derivative model proposed by the effective phase shift diffusion equation method [18] is

$$_tD_*^\alpha P'(\phi_{x'},\phi_{y'},\phi_{z'},t) = \left( D'_{f_2 x'x'} K'^{\beta_{x'}}_{x'}(t) \frac{\partial^{\beta_{x'}}}{\partial |x'|^{\beta_{x'}}} + D'_{f_2 y'y'} K'^{\beta_{y'}}_{y'}(t) \frac{\partial^{\beta_{y'}}}{\partial |y'|^{\beta_{y'}}} + D'_{f_2 z'z'} K'^{\beta_{z'}}_{z'}(t) \frac{\partial^{\beta_{z'}}}{\partial |z'|^{\beta_{z'}}} \right) \times P'(\phi_{x'},\phi_{y'},\phi_{z'},t) \quad (D.4)$$

which is hard to solve. By performing Fourier transform on both sides of Eq. (D.4), and setting $S(t) = P'(q_{x'}, q_{y'}, q_{z'}, t), q_{i'} = 1$ the following PFG signal attenuation equation can be obtained

$$_tD_*^\alpha S(t) = -\left( D'_{f_2 x'x'} |k'_{x'}(t)|^{\beta_{x'}} + D'_{f_2 y'y'} |k'_{y'}(t)|^{\beta_{y'}} + D'_{f_2 z'z'} |k'_{z'}(t)|^{\beta_{z'}} \right) S(t), \quad (D.5)$$

which is the same as Eq. (18).

II. Observing the signal intensity at the origin method:

The three-dimensional fractional diffusion equation can be written as

$$_tD_*^\alpha M'(x',y',z',t) = \left( D'_{f_2 x'x'} \frac{\partial^{\beta_{x'}}}{\partial |x'|^{\beta_{x'}}} + D'_{f_2 y'y'} \frac{\partial^{\beta_{y'}}}{\partial |y'|^{\beta_{y'}}} + D'_{f_2 z'z'} \frac{\partial^{\beta_{z'}}}{\partial |z'|^{\beta_{z'}}} \right) M'(x',y',z',t). \quad (D.6)$$

Substituting $M'(\mathbf{r}',t) = S(t)\exp(-i\mathbf{K}'(t)\cdot\mathbf{r}')$ into Eq. (D.6), yields

$$_tD_*^\alpha M'(\mathbf{r}',t) = -\left( D'_{f_2 x'x'} |k'_{x'}(t)|^{\beta_{x'}} + D'_{f_2 y'y'} |k'_{y'}(t)|^{\beta_{y'}} + D'_{f_2 z'z'} |k'_{z'}(t)|^{\beta_{z'}} \right) M'(\mathbf{r}',t). \quad (D.7)$$

At the origin where $\mathbf{r}' = 0$ and $i\gamma\mathbf{g}\cdot\mathbf{r}' = 0$, the gradient field has no effect on the magnetization phase and we have $M'(\mathbf{r}',t) = S(t)\exp(-i\mathbf{K}'(t)\cdot\mathbf{r}') = S(t)$, thus we get

$$_tD_*^\alpha S(t) = -\left( D'_{f_2 x'x'} |k'_{x'}(t)|^{\beta_{x'}} + D'_{f_2 y'y'} |k'_{y'}(t)|^{\beta_{y'}} + D'_{f_2 z'z'} |k'_{z'}(t)|^{\beta_{z'}} \right) S(t). \quad (D.8)$$

Eq. (D.8) reproduces Eqs. (18) and (D.5).

Only the fractional derivative model will be considered here. The fractal derivative model can be derived similarly.

**Acknowledgements**

The linguistic help from Thomas Caywood, Christian Fallen from writing center in Clark University, and Amoy Lin is acknowledged.